\documentstyle[preprint,aps]{revtex}
\newcommand{\be}{\begin{equation}} \newcommand{\ee}{\end{equation}}
\newcommand{\bea}{\begin{eqnarray}}\newcommand{\eea}{\end{eqnarray}}
\begin{document}
\draft
\title
{\LARGE Strangelets at Finite Temperature}
\author{M. G. Mustafa$^{ \ *}$ and  A. Ansari$^{ \ * }$}
\address
{Institute of Physics, Bhubaneswar-751005, India.}
\date{\today}
\maketitle
\begin{abstract}
\noindent   Within the MIT bag model picture of QCD, we have studied
the stability of finite size strangelets at finite temperature with
baryon number $A \leq 100$. The light
quarks $u$ and $d$ are considered massless while mass of the $s$ quark is
taken as $150$ MeV. Using discrete eigen energies of the non-interacting
quarks in the bag, grand canonical partition function is constructed and
then the free energy is minimized with respect to the bag radius. In
the $T\rightarrow 0$ limit clear shell structures are found which
persist upto about $T=10$ MeV. Infact for $A=6$ the shell structure
disappears only at $T > 20$ MeV.
\end{abstract}
\narrowtext

 \vfil
 \eject
\section{Introduction}
\label{sec:intro}
After Witten's conjecture in 1984 \cite{ew1} that the strange quark matter
(SQM) (consisting of roughly equal number of $u$, $d$ and $s$ quarks)
might be absolutely stable compared to iron, several model studies
have been made on its stability propertites at zero temperature
\cite{fj,bj,sq,jm,jm1,jm2,jm3,sc,gj} as
well as at finite temperatures \cite{rd,ca,sq1,sct,cg}. Particularly,
in view of relativistic
heavy ion collisions (RHIC) and the possibility of quark-gluon plasma
(QGP) formation, it has become interesting to investigate
theoretically as well as experimentally the possible existence of a stable or
metastable lump of SQM termed as strangelet
\cite{sq,jm,jm1,jm2,jm3,sc,gj}. For example, in Ref. [15]
it has been argued that when QGP starts cooling down and
hadronization begins $u$ and $d$ quarks combine with anti-strange
quarks ($\bar s$) producing $K^{+}$ and $K^0$ mesons and in the
process leaving enough number of $s$ quarks to condensate along with
$u$ and $d$ quarks into small strangelets at finite temperatures.
Hopefully the cooling is very fast so that the strangelets can really
be formed at sufficiently low temperatures, though finite.

Recently Gilson and Jaffe \cite{gj} have considered independent
single-particle (sp) shell model approach with quarks confined in a bag
with a given bag constant $B$. The energy of the bag for a given baryon
number $A$ is minimised with respect to the bag radius at zero
temperature. Even if the bulk energy per baryon $\epsilon_b \geq 950$
MeV, considering $u$ and $d$ massless and mass of $s$ quark $m_s \ =
\ 150$ MeV, they get shell structures leading to metastability against
nucleon decay. Madsen \cite{jm,jm3} has made a shell model versus liquid drop
model study for strangelets and gets metastability with life time
$\tau \geq 10^{-8}$ second. Boiling and evaporation of SQM at finite
temperature has also been studied in the literature \cite{rd,ca,sq1}.

Here we essentially want to extend to finite temperature  shell
model study of strangelets by Gilson and Jaffe \cite{gj} with $A \leq 100$.
That
is we want to study the thermodynamic properties of strangelets (big
MIT bag) with a scenario of thermalised hadronic phase after the
expansion and cooling of QGP in which the strangelets are embaded at
finite temperatures. To take care of the finite size of the strangelets
in the best way,
we construct a grand canonical partition function using the discrete
egien energies of quarks in the bag. As in most of the calculations
in the literature
we also consider $u$ and $d$ quarks to be massless and $m_s$ = 150
MeV. We do not have any restriction on the strangeness per baryon,
$i.e.$, $S/A$ ratio is unrestricted. Besidies, strangelets are
considered to be charge neutral and flavour eqilibrium is
maintained against weak decays \cite{rd}. We minimize the free energy for a
given $A$ as a function of bag radius $R$ at a given temperature and
look for the stable solutions. At present we are not constraining the
partition function to be colour-singlet. Since the considered
temperatures are rather low, $T\leq 40$ MeV the contributions from
gluons and anti-quarks are negligible. Also the temperature dependence
of the bag constant $B$ should not be important.

In the next sec. II we present briefly the formalism. In sec. II we
shall present our results and discussions along with a brief
conclusions in sec. IV.

\section{Formalism}
\label{sec:form}

We construct a small strangelet at finite temperature following
Ref. [10] as a gas of non-interacting fermions by filling the bag energy
states sequentially, obeying the Pauli exclusion principle. The
energy levels for quarks (massless $u$ and $d$-quarks and massive
$s$-quarks with mass, $m_s$ $=$ $150$ MeV) are obtained by solving the
Dirac equation for a spherical cavity of radius $R$ with linear
boundary condition \cite{dj}. The obtained sp eigen energies are
in the unit of $\hbar c/R$. Unlike in Ref. [10] we consider that there are
some electrons in the system to neutralize the electric charge, and the
flavour equilibrium is also maintained by weak interactions in which
neutrinos are produced. However even if the neutrinos are bound to the
system they
contribute very little to the energy and pressure of the system. Thus
we neglect them entirely. Electrons are also rather scarce but,
important for local charge neutrality of the strangelets. Of course, they
generate a coloumb barrier but of negligible importance which we will
be discussing
below.

In a statistical approach like Refs. \cite{am}, we construct
a grand canonical partition function in terms of participants
(quarks, anti-quarks, electrons and positrons) and study the
thermodynamics of
finite strangelets. We treat quarks and anti-quarks in the discrete
limit, and the electrons and positrons in the continuum limit.  In a
variational sense, for each $A$ we look for the minimum of the free
energy, $F(T,R)$ as function of bag radius $R$, at various values of
temperatures, $T$. At the minimum of $F(T,R)$ the quark pressure
balances the vacuum pressure, $B$ (the bag pressure constant) giving rise
to the mass of the strangelet as $4BV$, $V$ being the volume of the
system.

Normally in the bag model a zero-point energy is included as a
phenomenological term
of the form $-Z_0/R$, where fits to hadron spectra \cite{dj} indicates the
choice $Z_0 \ = \ 1.84$ for $B^{1/4} \ = \ 145$ MeV. Roughly half of
this phenomenological term has a physical basis in centre of mass
motion. For reasonable parameter values one finds a significant
effect of zero-point energy for $A < 10$, but the term quickly
becomes negligible for increasing $A$, where its contribution to the
energy per baryon goes like $A^{-4/3}$. Therefore, because of the great
uncertainty, like in Ref. [8],  zero-point energy is not included in our
calculation. The coloumb energy per baryon is negligible for most
strangelets due to the cancellation of $u$, $d$ and $s$-quark charges
and we ignore it \cite{gj}. We also ignore residual perturbative QCD interation
following Ref. [2] as for most of the purposes a non-zero value of
$\alpha_s$ can be absorbed in the decrease of $B$ provided
$m_s$ and bulk binding energy are held fixed.  Now, we briefly
outline the mathemtical steps involved in our calculation of hot
strangelets within the MIT Bag model.

The partition function for quarks and anti-quarks in
terms of single-particle energies \cite{am} is defined through
\be
\ln{{\cal Z}_{q({\bar q})}}  =  \sum_{\alpha} \Big \{ \sum_{q=u,d,s}
\ d^q_\alpha \ln \big ( 1 + e^{-\beta (\epsilon^q_\alpha - \mu_q)}
\big ) +  \sum_{{\bar q}={\bar u},{\bar d},{\bar s}}  \ d^{\bar
q}_\alpha \ln \big ( 1 + e^{-\beta (\epsilon^{\bar q}_\alpha -
\mu_{\bar q})} \big ) \Big \},  \label{qpat}
\ee
\noindent $\epsilon^{q({\bar q})}_\alpha$ are single particle energies
for quarks($q$) and anti-quarks(${\bar q}$), and
$\mu_q$ are chemical potential for different quark flavours
whereas $\mu_{\bar q} \ = \ - \mu_q$ are those for anti-quarks. $\beta \
= \ 1/T$, is inverse of the temperature. $d^{q({\bar q})}_\alpha$ is the
degeneracy with $3\times (2j+1)$ for each state $\alpha$ designated by
angular momentum $j$ and  $\kappa$, the Dirac quantum number \cite{dj}, and
$3$ is the colour factor for each flavour.

As mentioned earlier, we treat
electrons(positrons) in the continuum limit and their partition
function can be written as
\be
\ln{{\cal Z}_e} = {gV\over {2\pi^2}} \left [ \int {\epsilon^2_{e}}  \ln
\left ( 1+e^{-\beta (\epsilon_{e} -\mu_{e})} \right ) {\rm
d}\epsilon_{e} + \int
{\epsilon^2_{e}} \ln \left ( 1+e^{-\beta (\epsilon_{e} + \mu_{e})}
\right ) {\rm d}\epsilon_{e} \right ] , \label{epat}
\ee
\noindent where $g \ = \ 2$ is the spin degeneracy for $e^-(e^+)$ and
$\mu_{e}$ is the chemical potential for $e^-$ and $- \mu_{e}$ is that
for $e^+$. However, one can neglect the $e^+$ as it contributes
nothing in the temperature range we will be working. Infact,
contributions of $e^-$ are also very small.

We shall assume that the weak interaction maintains equilibrium
between the different quark flavours through the processes:
\bea
d \ \longrightarrow \ u \ + \ e^- \ + \ {\bar \nu}_{e} \ , && s \
\longrightarrow \  u \ + \ e^- \ + \ {\bar \nu}_{e} \ , \nonumber \\
u \  + \ e^- \ \longleftrightarrow \ d \ +  \ {\nu}_{e} \ , &&
u \  + \ e^- \ \longleftrightarrow \ s \ +  \ {\nu}_{e} \ , \nonumber \\
u \  + \ s \ &\longleftrightarrow & \ d \ +  \ u . \ \label{ew}
\eea
\noindent However the contributions of electrons  to the energy and
pressure in the study of equilibrium properties of the strangelets are
neglected.

\noindent The dynamical chemical equilibrium among the participants yeilds:
\be
\mu_d \ = \ \mu_s \ = \ \mu  \ , {\rm {and}} \ \mu_u \ = \ \mu \ - \
\mu_{e} \ . \label{che}
\ee

\noindent The local charge neutrality gives a relation among number densities :
\bea
{2\over 3} \left ( n_u - n_{\bar u} \right ) \ - \ {1\over 3} \left (
n_d - n_{\bar d} \right ) \ & - & \ {1\over 3} \left ( n_s - n_{\bar
s} \right )  \ - \ \left ( n_{e^-} - n_{e^+} \right ) \ = \ 0 \ ,
\nonumber \\
2\triangle n_u \ - \ \triangle n_d \ - \ &\triangle n_s & \ - \
3\triangle n_e \ = \ 0 \ . \label{nd}
\eea
\noindent The number densities are given as
\be
n_i \ = \ {N_i\over V} \ = \ {T\over V} {\partial \over {\partial \mu_i}}
\left ( \ln{{\cal Z}_i}\right ) \ \ , \label{gnd}
\ee
\noindent where $V$ is the volume of the system, $N_i$ is the number
of each participants, and $i= q$, ${\bar q}$ and $e$. The change in number
densities are given, respectively, as
\be
\triangle n_q = {1\over V} \sum_{\alpha} \left [ \sum_q {d^q_{\alpha}
\over {\left (e^{\beta (\epsilon^q_{\alpha} -\mu_q)} + 1 \right ) }}
- \sum_{\bar q} {d^{\bar q}_{\alpha} \over {\left
(e^{\beta (\epsilon^{\bar q}_{\alpha} + \mu_q)} + 1 \right ) }}
\right ] , \label{cndq}
\ee
\noindent and,
\be
\triangle n_e = {g\over {2\pi^2}} \int {\rm d}\epsilon_e
\left [  {1
\over {\left (e^{\beta (\epsilon_{e} -\mu_e)} + 1 \right ) }}
-  {1 \over {\left
(e^{\beta (\epsilon_{e} + \mu_e)} + 1 \right ) }}
\right ]  \label{cnde}
\ee

A baryon number $A$ is imposed by adjusting quark chemical potentials
such that the excess number of $q$ over ${\bar q}$ is $3A$, namely,
\be
\triangle N_q = N_q  -  N_{\bar q}  =  (N_u  -  N_{\bar u}) + (N_d  -
N_{\bar d}) + (N_s -  N_{\bar s}) = 3 A \ . \label{qn}
\ee
\noindent The energy and free energy of the whole system can be
written, respectively, as
\be
E (T, R) \ = \ T^2{\partial \over {\partial T}}
\left (\ln{{\cal Z}_i}\right ) \ + \ \mu_q
\triangle N_q \ + \ BV  \ \ . \label{e}
\ee
\noindent and,
\be
F ( T, R) \ = \ - T\ln{{\cal Z}_i} \ + \ \mu_q
\triangle N_q \ + \ BV  \ \ . \label{f}
\ee
\noindent where $BV$ is the bag volume energy term \cite{dj}.

The pressure generated by the participants gas
\be
P \ = \ - \left ( {\partial \over {\partial V}} F(T,R) \right
)_{T,{\triangle}N_q} \ \ , \label{p}
\ee
\noindent is balanced by the bag pressure constant, $B$ which in turn
gives the stability condition of the system. Then, the equilibrium
energy of the system as given by (\ref{e}) is
\be
E(T,R) \ = \ 4B V  \ \ \ , \label{ee}
\ee
\noindent where $\ T^2{\partial \over {\partial T}}
\left ( \ln{{\cal Z}_i} \right ) \ + \ \mu_q
\triangle N_q \ = \ 3BV$ .

\section{Result and Discussion}
\label{sec:res}

The physical behaviour of a system at a temperature $T$ is governed
by the properties of its free energy. Treating the bag like a
many-body system at a temperature $T$, its stability features are
studied with the variation of free energy, $F(T,R)$, considering $R$
as a variational parameter. It may be reminded that the
single-particle energies of quarks are in the units of ${\hbar c/R}$
and through this $R$ enters in, for instance, eq.(\ref{qpat}). For a
fixed value of $T$, $A$ and $\mu_u$ ($u$-quark chemical potential),
we assume a trial $R$. Then we solve
(\ref{che}) and (\ref{nd}) simultaneously for $\mu$ ($d$ and
$s$-quark chemical potential) and $\mu_e$ (electron chemical
potential). After this we calculate $A$ from number constraint
(\ref{qn}) and compare with the chosen value of $A$. The above
procedure is repeated changing $\mu_u$ till baryon number constraint
(\ref{qn}) is satisfied upto certain accuracy with the chosen value
of $A$. Then for that given $T$, same $A$ and trial $R$ we calculate
$F(T,R)$ of the strangelet using (\ref{f}). After this with those $T$
and $A$, the whole procedure is again repeated  changing $R$
(keeping it in mind that each time (\ref{che}) and (\ref{nd})
simultaneously and (\ref{qn}) separately have to be satisfied) until
the $F(T,R)$ is minimum obeying the stability condition (\ref{p}).
Now we
have a equlibrium value of radius $R$ and energy $E(T,R)$ of a
strangelet for a given $T$ and $A$. Then for a fixed $T$ one can find
the stability of strangelets for different values of $A$.

In Fig. 1 we have plotted the energy per baryon $\epsilon = E/A$ for
strangelets as a function of $A$ (upto $100$) for $B^{1/4}=145$ MeV
at equilibrium (\ref{p}), for various $T$($0.5$,
$10$, $20$, $30$, $40$ MeV). The curve for $T$ = 0.5 MeV basically
represents the zero-temperature results of Gilson and Jaffe \cite{gj}. Here,
we have chosen  $B^{1/4}=145$ MeV to have the SQM
absolutely stable (compared to nucleons) in the bulk phase and we
have the bulk energy $\epsilon_b \ = \ 850$ MeV. The
metastability for strangelets and various decay modes have been
studied by Gilson and Jaffe \cite{gj} in four different parameter space
within the MIT bag model. The strangelets considered here are stable
against neutron decay for $A \geq 10$, where $(m_n-\epsilon)$ is the
binding energy (relative to neutron mass $m_n \ \approx \ 940$ MeV)
of the strangelets. With the choice of higher values of $B$ these curves
will essentially move up.

Now, one can notice in Fig. 1 that $E/A$ increases with increase of
$T$ and until $T=30$ MeV strangelets ($A \geq 20$) are stable compared
to nucleons $i.e.$, $E/A < 940$ MeV. At $T > 35$ MeV, $E/A > 940$
implying there is no stable configurations and the strangelets become
unstable compared to nucleons. In the zero temperature
limit (due to numerical problems we take $T= 0.5$ MeV) the $E/A$
approaches a bluk limit at $A\approx 100$, whereas energy grows
significantly for low
$A$. For $s$-quark mass, as expected, shell closures appear for $A \
=$ $6, \ 18, \ 24, \ 42, \ 54, \ 84, \ 102, \ 150, \cdots $ (of course
$102$ and $150$ are not shown in Fig. 1). At every shell closure
$E/A$ has a noticeable dip  which corresponds to a more stable
strangelet compared to the stangelets in the neighbourhood. With
higher values of $B$ these shell closures will give rise to
metastable strangelets. Also their positions may somewhat change with
change of $B$ and $m_s$.
Among these shell closures with low mass strange quark
($m_s = 150$ MeV), $A=6$ is the most significant shell closure whereas
those with
higher $A$ are less conspicuous ones. Our results at $T=0.5$ MeV more
or less similar to those of Gilson and Jaffe \cite{gj} at $T$ = 0. Our
calculation differs from that of Gilson and Jaffe \cite{gj} as we consider
charge
neutrality and chemical equilibrium and ignore zero-point energy
correction. The most important point is that we are able to see the
effect of temperature on shell effects and the latter persists upto
about $T$ = 10 MeV. Infact, at $A$ = 6 the shell effect persists even
upto $T$ = 20 MeV. Though for higher values of $B$ these states
become metastable, the shell features remain unaffected and we have
checked it for $B^{1/4}$ = 160, 175 MeV.

Now we can analyze how bag states are filled up by quarks. The dynamics
is as follows. At $T\rightarrow 0$ (of course here $T=0.5$ MeV) and  $A$
upto $3$ the fermi surface is such that it only
accomodates non-strange $1s_{1/2}$ states and strange $1s_{1/2}$ state
(corresponding to massive $s$-quark)
lies outside the fermi surface created by non-strange quarks. At
$A=3$, non-strange $1s_{1/2}$ state saturates with $6$ $d$-quarks,
and is half-filled with $3$ $u$-quarks and
strange $1s_{1/2}$ state remains empty. This can be seen clearly from
Figs. 2 and 3. Upto $A=3$ there is no strangelets rather they are
$ud$-droplets and their $E/A \geq 940$ MeV (see Fig. 1). In the
region $3<A<6$, non-strange $1s_{1/2}$ state for $d$-quarks can not
accomodate further as it is already saturated (see Fig. 2). Same
feature can
also be seen from Fig. 3 where $d$-fraction is decreasing with
increasing $A$ in this region ($3<A<6$). The non-strange $1s_{1/2}$
level for $u$-quarks fills up with constant rate as $u$-fraction
remains constant here and throughout (see Fig. 3). The strange
$1s_{1/2}$ level starts accomodating because, now, the strange quark
mass is lower than the fermi energy generated by $ud$-droplets. So,
the opening of a new flavour degree of freedom tends to lower the
fermi energy and hence the mass of the strangelets (see Fig. 1). At
$A=6$, $1s_{1/2}$ states for all the three $u$, $d$ and $s$-quarks
are completely filled and first shell closure occures.
This can be seen clearly from Figs. 2 and 3.

After one shell closure, it again becomes favourable to add
non-strange quarks to the system. This continues until $d$-quarks
have
fully occupied the corresponding non-strange shell. Then again
strange levels become energitically favourable to fill them up. When
the corresponding strange level
saturates, the next shell closure occures. So we find that
non-strange and strange states in MIT bag are filled up sequentially
obeying Pauli exclusion principle and every shell closure occures
due to the completion of a particular strange level (see Figs.
2 and 3).

Now, we discuss how charge neutrality (\ref{nd}) is satisfied in
$T\rightarrow 0$ limit ( here $T=0.5$ MeV). Above we have already
discussed how levels
are filled up in the bag for strangelets. When only non-strange
states ($u$, $d$ states) are filling up (see Figs. 2 and 3) with
particular $j$ and $\kappa$, corresponding strange state is empty and
Fig. 4 indicates that the number of electron is zero. This implies
that the charge neutrality (\ref{nd}) is only satisfied among
different quark flavours of filled and currently filling states. This
situation continues unless $d$-state is saturated with full occupancy
and $u$-state is just half-filled. Soon after this the half-filled
$u$-state starts filling further and corresponding empty $s$-state
starts accomodating particles (see Figs. 2 and 3). The charge
neutrality (\ref{nd}) demands electrons until $u$ and $s$-states are
completely occupied. This feature can be seen from Fig.4.  When low
lying $u$, $d$ and $s$-states
are completely filled, a shell closure occures $i.e.$, there is equal
number of $u$, $d$ and $s$-quarks and charge neutrality implies no
electrons. Immediately after a shell closure, feature is same till
next shell closure appears because first to fill is next non-strange
shell and then strange shell. We can see from Figs. 2, 3 and 4 that
this situation recurs atleast upto $A=100$.
The strangelets with closed shells are relatively more stable and
less chemically reactive because the decay of an $s$ quark to $d$
quark would be suppressed or forbidden because lowest single-particle
levels are fully occupied.
We find from Fig. 3 that in the temperature region $20<T<30$, the
$s$-fraction is around $0.85$ and that of $d$ $\sim$ $1.15$. However,
$u$-fraction remains constant throughout at unity at all $T$. Fig.
4 shows that with the increase of temperature the electron content of the
strangelet increases. The increase in the number of electrons
indicates higher neutrino emissivity from strangelets. It also
improves the heat transport across the bag surface in a system such as a
neutron star with the SQM core.

We find that at high temperature ($T>35$) where
strangelets become unstable these are not in flavour equilibrium. So
strangelets can decay via weak semileptonic decays, weak radiative
decays and  electron capture all of which can have $\triangle A = 0$.
This is also applicable even at $T\rightarrow 0$ limit. At $T=0$,
$E/A$ determines whether the system is stable against decay into
neutron or nuclei. For $E/A < 930$ MeV, strangelets can not decay
even into $^{56}{\rm Fe}$ nuclei. As long as $E/A < 930$ MeV, emission
of an $\alpha$-particle goes via energy fluctuations but only subject
to decay modes which reduce $A$. For higher temperature, $E/A > 930$
MeV, these fluctuations are not needed and the process can proceed
faster. At higher temperature neutron decay might take place if the
difference of energy between two strangelets is greater than $m_n$,
$\triangle S = 0(-1)$ and $\triangle A = -1$. Simililarly, for
the $\Lambda$, $\Sigma$, $\Xi$, $\Omega$ decays, one should have
$\triangle A = -1$ and $\triangle S = -1, \ -1, \ -2, \ -3$,
respectively.

\section{Conclusion}
\label{sec:con}

Considering non-interacting massless $u$ and $d$-quarks and massive
$s$-quarks ($m_s$=150 MeV) confined in an MIT bag at finite temperature
thermodynamic properties of finite size strangelets with baryon number
$A\leq 100$ have been studied following the method of quantum statistics.
As discussed in the text, the grand canonical partition function at $T=0.5$
MeV describes rather well the $T\rightarrow 0$ limit properties of the
strangelets showing shell structures and the possibility of stable or
metastable strangelets depending on the employed value of the bag constant
$B$. As far we are aware, this is seen for the first time that the shell
structure gets washed away (no possibility of metastable states) only at
$T\approx 10$ MeV. Infact for $A=6$ the shell structure vanishes only at
$T\geq 20$ MeV. We find that for $B^{1/4}=145$ MeV which is a physically
accepted value, the strangelets remain stable against nucleon decay till
$T=30$ MeV.

Next we are planning to repeat this calculation with the use of a
colour-singlet partition function which would incorporate the most
important aspect of QCD interaction.

\vfil
\eject
\begin{figure}
\caption{ Energy per baryon ($E/A$) as a function of baryon
number $A$ for different temperatures ($T=$0.5, 10, 20, 30, 40 MeV)
with $B^{1/4}=145$ MeV and $s-$quark mass $m_s$=150 MeV.}
\end{figure}
\begin{figure}
\caption{Number of quarks ($d$ and $s-$quarks) as a function
of baryon
number $A$ for different temperatures ($T=$0.5, 10 MeV)
with $B^{1/4}=145$ MeV and $s-$quark mass $m_s$=150 MeV.}
\end{figure}
\begin{figure}
\caption{Quark-fraction as a function of baryon
number $A$ for different temperatures ($T=$0.5, 10, 20, 30 MeV)
with $B^{1/4}=145$ MeV and $s-$quark mass $m_s$=150 MeV.}
\end{figure}
\begin{figure}
\caption{ Number of electrons  as a function of baryon
number $A$ for different temperatures ($T=$0.5, 10, 20, 30 MeV)
with $B^{1/4}=145$ MeV and $s-$quark mass $m_s$=150 MeV.}
\end{figure}
\vfil
\eject
\end{document}